\documentclass[twocolumn,showpacs,preprintnumbers,amsmath,amssymb,superscriptaddress]{revtex4}
\usepackage{graphicx}
\usepackage{dcolumn}
\usepackage{bm}
\usepackage{color}

\usepackage[textsize=footnotesize]{todonotes}

\DeclareMathOperator{\erfc}{erfc}

\renewcommand\vec{\mathbf}
\renewcommand{\(}{\left(}
\renewcommand{\)}{\right)}

\def \equi#1{\mathrel{\mathop{\kern 0pt\sim}\limits_{#1}}} 

\begin{document}

\title{$N$-tag Probability Law of the  Symmetric Exclusion  Process}

\date{\today}
\author{Alexis Poncet}
\affiliation{LPTMC, CNRS/Sorbonne Universit\'e, 4 Place Jussieu, F-75005 Paris, France}
\author{Olivier B\'enichou}
\affiliation{LPTMC, CNRS/Sorbonne Universit\'e, 4 Place Jussieu, F-75005 Paris, France}
\author{Vincent D\'emery}
\affiliation{Gulliver, CNRS, ESPCI Paris, PSL Research University, 10 rue Vauquelin, Paris, France}
\author{Gleb Oshanin}
\affiliation{LPTMC, CNRS/Sorbonne Universit\'e, 4 Place Jussieu, F-75005 Paris, France}

\begin{abstract}
The Symmetric Exclusion Process (SEP), in which particles  hop symmetrically on a discrete line with hard-core constraints, is a paradigmatic model of subdiffusion in confined systems. This  anomalous behavior  is a direct consequence of strong spatial correlations induced by the requirement that the particles cannot overtake each other. Even if this  fact has been recognised qualitatively for a long time, up to now there is no full quantitative determination of these correlations. Here we study the joint probability distribution of an arbitrary number of tagged particles in the SEP. We determine analytically  the large time limit of all cumulants  for an arbitrary density of particles, and their  full dynamics in the high density limit. In this limit, we unveil  a universal scaling form shared by the cumulants and
 obtain the time-dependent large deviation function of the problem.
\end{abstract}

\maketitle

{\it Introduction.} Single-file diffusion refers to the motion of particles in  narrow channels, in which the geometrical constraints do not permit the particles to bypass each other. The very fact that the initial order is maintained at all times   leads to a subdiffusive behavior $\langle X_t^2\rangle \propto\sqrt{t}$ of the position of any tagged particle (TP) \cite{Harris:1965}, as opposed to the regular  diffusion scaling $\langle X_t^2\rangle \propto{t}$. This theoretical prediction has been experimentally observed by microrheology in zeolites, transport of confined colloidal particles, or dipolar spheres in circular channels \cite{Gupta:1995,Hahn:1996,Wei:2000,Meersmann:2000,Lin:2005}.

A minimal model of  single file diffusion is the symmetric exclusion process (SEP). Here,  particles, present at a density $\rho$, perform symmetric continuous time random walks on a one dimensional lattice with unit jump rate, and hard-core exclusion is enforced by allowing at most one particle per site. A key result is that the long time behavior of the variance of the position of a TP initially located at the origin obeys:
\begin{equation}
\langle X_t^2\rangle \equi{t\to\infty} \frac{1-\rho}{\rho}\sqrt{\frac{2t}{\pi}}. 
\end{equation}
The SEP has now become a paradigmatic model of subdiffusion in confined systems  and it has generated a huge number of works in the mathematical and physical literature (see, e.g., Refs.~\cite{Levitt:1973,Fedders:1978,Alexander:1978,Arratia:1983,Lizana:2010,Taloni:2008,Gradenigo:2012}). 
 Recent advances include the calculation of the cumulants of $X_t$ in the dense limit $\rho\to1$ \cite{Illien:2013} or at long time  for any density \cite{Imamura:2017}. 
While the SEP in its original formulation provides a model of subdiffusion in crowded equilibrium systems, important extensions to non-equilibrium situations have recently been considered. In Ref.~\cite{Imamura:2017}, all the cumulants of a symmetric TP immersed in a step initial profile with different densities  of particles on the left and on the right of the TP are calculated. In the other intrinsically out of equilibrium situation of a driven TP in a SEP, the mean position~\cite{Burlatsky:1996b,Landim:1998a} and all higher order moments in the dense limit~\cite{Illien:2013} have been calculated, and shown to grow anomalously like $\sqrt{t}$.  

This collection of anomalous behaviors in the SEP is a direct consequence of strong spatial correlations in the single file geometry. Even if this  fact has been recognised qualitatively for a long time, up to now there is no full quantitative determination of these correlations. As a matter of fact, all the results mentioned above concern observables associated with a {\it single} TP. A complete characterisation of the correlations  requires the knowledge of {\it several} TP observables.  
To date, the only  available results  concern the case of two  TPs.   Two-point correlation functions have  been analysed  either by using a stochastic  harmonic theory  (Edwards-Wilkinson dynamics) \cite{Majumdar:1991} or for the so-called random average process \cite{Rajesh:2001,Cividini:2016}, which displays several  qualitative features similar to the dilute limit of the SEP.  In the continuous space description, which can be seen as the dilute limit $\rho\to0$ of the SEP, a two-tag probability distribution has been determined \cite{Sabhapandit:2015}.

Here we study the full joint distribution of an arbitrary number $N$ of TPs in the SEP. More precisely, we determine (i)  the large time limit of all cumulants  for an arbitrary density of particles and (ii) their  full dynamics in the dense limit $\rho\to1$. This last result permits us to unveil  a universal scaling form shared by all cumulants and
to obtain the time dependent large deviation function of the problem.

\begin{figure}
 \includegraphics[width=8cm]{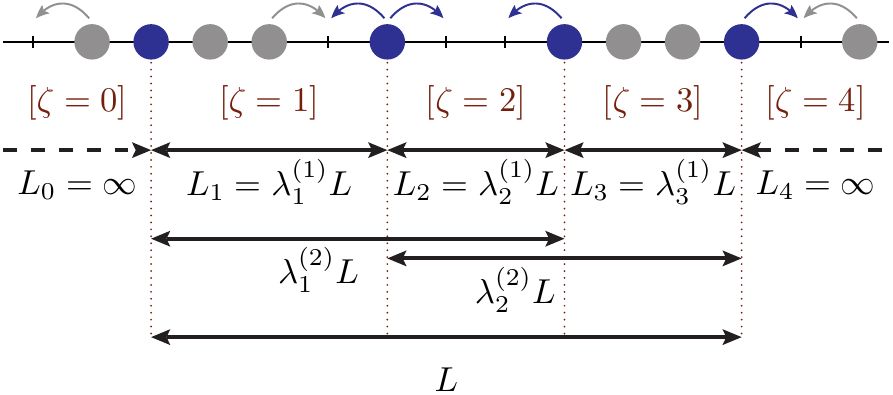}
 \caption{Summary of our notations in the case of $N=4$ TPs.
 The blue particles are the TPs, the gray ones are the other particles. The curved
 arrows show the allowed moves.}
 \label{fig1}
\end{figure}

{\it Model.} Let us consider hard-core particles on a discrete one-dimensional line.
The mean density of particles is denoted by $\rho$.
The particles follow  symmetric random walks with hard-core exclusion. In order to characterise the correlation functions involved in this system, we tag $N$ of these particles, the TPs.
The initial distances between them are denoted by $L_1, \dots, L_{N-1}$ and the initial position of the  $i$-th TP is $X_i^0 = \sum_{j=1}^{i-1} L_j$ for $i\ge2$ and $X_1^0 =0$ (Fig.~\ref{fig1}).

{\it Large time behavior at any density.} We start by showing that the long time behavior of the cumulants can be determined from general arguments, valid at any density $\rho$. 
This relies on the fact that  the distance between two neighboring TPs reaches an equilibrium distribution,  that can be obtained from the following arguments.
When two neighboring TPs are initially at a distance $L$,  each of the $L-1$ sites between them is occupied with probability $\rho$: the number of bath particles between the TPs, denoted $k$, thus follows a binomial distribution of parameters $L-1$ and $\rho$.
At equilibrium, the number of vacancies between these two TPs follows a negative binomial distribution: it is the law of the number of failures before $k+1$ successes when the probability of a success is $\rho$.
Finally, the distance $\Delta$ between the two TPs is given by the sum of the number of bath particles and vacancies between them. It thus reaches a stationary value  and follows the distribution
\begin{equation} \label{eq:lawDist}
P_\Delta(\delta) = \sum_{k=0}^{L-1}\binom{L-1}{k}\binom{\delta-1}{k} \rho^{2k+1}(1-\rho)^{L+\delta-2k-2}.
\end{equation}
We find a very good agreement of this law with numerical simulations~\cite{SupplMat}.
We now note that the cumulants $\kappa^{(N)}_{p_1,\dots,p_N}(t)$ involving several TPs (see Eq.~\eqref{eq:defCum} for a precise definition)
 can be written as a sum of moments involving a single TP and moments involving distances.  For instance, $\kappa^{(2)}_{11}(t)= \langle X_1(t)X_2(t) \rangle=\langle X_1(t)^2 \rangle+\langle X_1(t)[X_2(t)-X_1(t)] \rangle$. 
The crucial point is that $\langle X_1(t)^2 \rangle\sim \sqrt{t}$, while $\langle (X_2(t)-X_1(t))^2 \rangle ={\mathcal O} (t^0)$, which implies   from the Cauchy-Schwarz theorem that $\langle X_1(t)[X_2(t)-X_1(t)] \rangle=  {\mathcal O}\left(t^{1/4}\right)$. 
The large time behavior of $\kappa^{(2)}_{11}(t)$ is thus given by the large time behavior of the single TP cumulant of the same order, $\kappa^{(1)}_2(t) = \langle X_1(t)^2 \rangle$.
This is true for all the cumulants, leading finally to 
\begin{equation} \label{eq:equal1N}
\lim_{t\to\infty}\frac{\kappa^{(N)}_{p_1, \dots, p_N}}{\sqrt{t}}
= \lim_{t\to\infty}\frac{\kappa^{(1)}_{p_1 + \dots + p_N}}{\sqrt{t}}
= B_{p_1+\dots+p_N},
\end{equation}
where the constants $B_{k}$, involved in the  single tagged particle problem, have been determined in Ref. \cite{Imamura:2017}.
Equation \eqref{eq:equal1N} implies that  the group of $N$ TPs behaves at long times like a single TP.
However,  the approach to this asymptotic state remains unknown. 
Below, we determine completely the dynamics of the cumulants in the limit of a dense system, which constitutes the core of this Letter.

{\it Dense limit.} Following Refs. \cite{Brummelhuis:1989a,Benichou:2002qq,Illien:2013,Benichou:2013d,Illien:2014,PhysRevLett.115.220601}, we  focus on the limit of a dense system ($\rho\to 1$) and  follow the evolution of the vacancies, rather than the particles, in a discrete time. We assume that at each time step, each  vacancy is moved  to one of  its nearest neighbour sites, with equal probability. 
It thus performs a symmetrical nearest neighbor  random walk. Note that  a complete description of the dynamics would require  additional rules for cases where two vacancies are adjacent or have common neighbours; however, these cases contribute only to ${\mathcal O}((1-\rho)^2)$, and can thus be left  unstated.
 In this limit, one can approximate the motion of the TPs as being
generated by the vacancies interacting independently with them: in the large density limit the events
corresponding to two vacancies interacting simultaneously with the TP happen with negligible probability. These rules are the discrete counterpart of the continuous time version of the SEP described above in the dense limit, as shown in  \cite{Benichou:2002qq,Illien:2013}. They allow us to obtain the  dynamics of the SEP, in the scaling regime defined below.

We start from a system of size $\mathcal N$ with $M$ vacancies and denote
$\vec Y(t) = (X_i(t)-X_i^0)_{i=1}^N$
the vector of the displacements of the $N$ TPs.
 In the dense limit $\rho\to 1$, the contributions due to the vacancies can be summed, so that we can
link (i) the probability $P^{(t)} (\vec Y|\{Z_j\})$ of having displacements $\vec Y$ at
time $t$ knowing that the $M$ vacancies started at sites $Z_1\dots Z_M$
to (ii) the probability $p^{(t)}_Z(\vec Y)$ that the displacements of the TPs are $\vec Y$ at time $t$
due to a {\it single} vacancy that was initially at site $Z$~\cite{Brummelhuis:1989a,Benichou:2002qq} by:
\begin{equation}\label{eq:approx}
 P^{(t)} (\vec Y|\{Z_j\}) \underset{\rho\to 1}{\sim}
  \sum_{\vec Y_1,\dots,\vec Y_M}
  \delta_{\vec Y, \vec Y_1+\dots+\vec Y_M}
  \prod_{j=1}^M p^{(t)}_{Z_j}(\vec Y_j)
\end{equation}
Taking the Fourier transform with respect to $\vec Y$,
averaging over the initial positions of the vacancies
and finally  taking the thermodynamic limit   $\mathcal{N},M\to\infty$ with $\rho_0 \equiv 1-\rho = M / \mathcal{N}$ remaining constant,
the second characteristic  function 
\begin{equation}
\psi^{(t)}(\vec k) \equiv 
\ln \left[\left\langle \sum_{\vec Y \in \mathbb{Z}^N}P^{(t)} (\vec Y|\{Z_j\})e^{i\vec k\cdot \vec Y}\right\rangle_{\{Z_j\}}\right]
\end{equation}
is found to be given by  
\begin{equation}
\lim_{\rho_0\to 0}\frac{\psi^{(t)}(\vec k)}{\rho_0} =
\sum_{Z \neq 0,L} \tilde q_Z^{(t)}(\vec k) \label{eq:psi1}
\end{equation}
where $\tilde q_Z^{(t)}(\vec k)\equiv\tilde p_Z^{(t)}(\vec k) - 1$, $p_Z^{(t)}(\vec k)$
being the Fourier transform of $p_Z^{(t)}(\vec Y)$.
By definition, $\psi$ gives the $N$-tag cumulant of the displacements with coefficients
$p_1,\dots, p_N$ as:
\begin{equation} \label{eq:defCum}
 \kappa^{(N)}_{p_1,\dots,p_N} = (-i)^{p_1+\dots+p_N} \left.
 \frac{\partial^{p_1+\dots+p_N}\psi}{\partial k_1^{p_1}\dots \partial k_N^{p_N}} 
 \right|_{\vec k = \vec 0}.
\end{equation}

The next step of the calculation consists in determining the single-vacancy probability  $\tilde q^{(t)}_Z(\vec k)$ involved in Eq.~\eqref{eq:psi1}, by considering   a system containing a single vacancy.
An intrinsic technical difficulty in a problem with several TPs is that the distance between them is not constant, which in turn makes the first-passage properties of the vacancy to the TPs time-dependent.
However, this difficulty can be overcome in the  one dimensional situation considered here because, in the case  of a single vacancy, the distances between TPs can only assume two values
depending on the initial position $Z$ of the vacancy: we define
$\zeta(Z) = i$ if the vacancy starts between TP $i$ and TP $i+1$,
and $\zeta(Z) = 0$ (resp. $\zeta(Z) = N$)
if it starts on the left (resp. on the right) (Fig. \ref{fig1}).
The distance to be considered when the vacancy is, at some instant,
between TP $i$ and TP $i+1$ is then given by $L_i^{(\zeta)} = L_i + 1$ if $\zeta\neq i$
or $L_i^{(i)} = L_i$ if $\zeta = i$.

The key to obtain $\tilde q_Z^{(t)}(\vec k)$  is to introduce the  first-passage probability 
$F^{(t)}_{\eta, Z}$ that 
 the vacancy that started from site $Z$ at time
$0$ arrives for the first time to the position of one of the TPs at time $t$, conditioned
by the fact that it was on the ``adjacent site'' $\eta$ at time $t-1$.
The adjacent site $\eta = i$ (resp. $\eta = -i$) is defined as the site to the right 
(resp. left) of the $i$-th TP.
In analogy with $q_Z$ and $F_{\eta, Z}$, we introduce quantities related to an adjacent site
$\nu$: $q_\nu^{(t, \zeta)}$ and $F_{\eta,\nu}^{(t, \zeta)}$ (that depend on the distances
between TPs, thus on $\zeta$).
One can now partition over the first passage of the vacancy
to the site of one of the TPs to get an expression for $\tilde q_Z$~\cite{SupplMat}.
\begin{equation} \label{eq:qAll}
 \tilde q_Z^{(t)}(\vec k) = - \sum_{j=0}^t \sum_{\nu}
 \left[1 - \(1+\tilde q_{-\nu}^{(t-j,\zeta(Z))} (\vec k)\) F^{(j)}_{\nu, Z}\right].
\end{equation}

To obtain $\tilde q^{(t, \zeta)}_\eta$, we decompose the propagator of the displacements over the successive passages of the vacancy to
the position of one of the TPs
using $\vec e_{\pm 1} = (\pm 1, 0, \dots)$, \dots, $\vec e_{\pm N} = (0, \dots, 0, \pm 1)$
as a basis for the displacements $\vec Y$~\cite{SupplMat}.
This writes as a time convolution of quantities $F_{\nu,\eta}^{(t, \zeta)}$:
the Laplace transform,
$\hat{\tilde q}^{(\zeta)}_\eta (\vec k, \xi) = \sum_{t=0}^\infty \xi^t \tilde q^{(t, \zeta)}_\eta (\vec k)$,
writes as an infinite sum of matrix powers, giving:
\begin{multline} \label{eq:qBorder}
 \hat{\tilde q}^{(\zeta)}_\eta (\vec k, \xi) = \frac{1}{1-\xi} \sum_{\mu,\nu}
 \{[1-T^{(\zeta)}(\vec k, \xi)]^{-1}\}_{\nu\mu} \\
 \times \(1-e^{-i\,\vec k\cdot\vec e_\nu}\) e^{i\,\vec k\cdot\vec e_\mu}
 \hat F^{(\zeta)}_{\mu\eta}(\xi).
\end{multline}
The matrix $T$ is defined by $T^{(\zeta)}(\vec k,\xi)_{\nu\mu} = \hat F^{(\zeta)}_{\nu, -\mu} (\xi) e^{i\,\vec k\cdot\vec e_{\nu}}$.

Introducing Eq.~(\ref{eq:qAll}) into Eq.~(\ref{eq:psi1}), one gets an expression for the Laplace transform
of the second characteristic function:
\begin{align} \label{eq:psiSol}
 \lim_{\rho_0\to 0}\frac{\hat \psi (\vec k, \xi)}{\rho_0} =& - \sum_\nu \bigg[
 \frac{1}{1-\xi} \nonumber \\
 &- \(\frac{1}{1-\xi} + \hat{\tilde q}^{(\zeta)}_{-\nu} (\vec k, \xi)\)
 e^{i\,\vec k\vec e_\nu} \bigg] h_\zeta(\xi) \\
 h_\zeta(\xi) =& \sum_{Z\notin \{X_i^0\}} \hat F_{\zeta,Z} (\xi).
 \end{align}
Here we defined $\zeta = \zeta(\nu) \in [0,N]$ by $\zeta = \nu$ if $\nu > 0$ and
$\zeta = -\nu-1$ if $\nu < 0$.

Finally, the quantities that we need are the probabilities to go from one adjacent site to another:
$\hat F_{-1,-1} = \hat F_{+N, +N}$,
$\hat F_{\mu,\mu}^{(\zeta)} = \hat F_{-(\mu+1),-(\mu+1)}^{(\zeta)}$,
$\hat F_{\mu,-(\mu+1)}^{(\zeta)} = \hat F_{-(\mu+1),\mu}^{(\zeta)}$,
and the sums $h_0 = h_N$ and $h_\mu$ ($\mu = 1,\dots, N-1$).
The other $\hat F_{\nu, \eta}$ are zero.
These quantities can be computed explicitly  using classical results on first-passage times of symmetric one dimensional random walks \cite{Hughes:1995}, which completes the determination of the second characteristic function (see \cite{SupplMat} for an explicit expression).

\begin{figure*}
 \includegraphics[width=17cm]{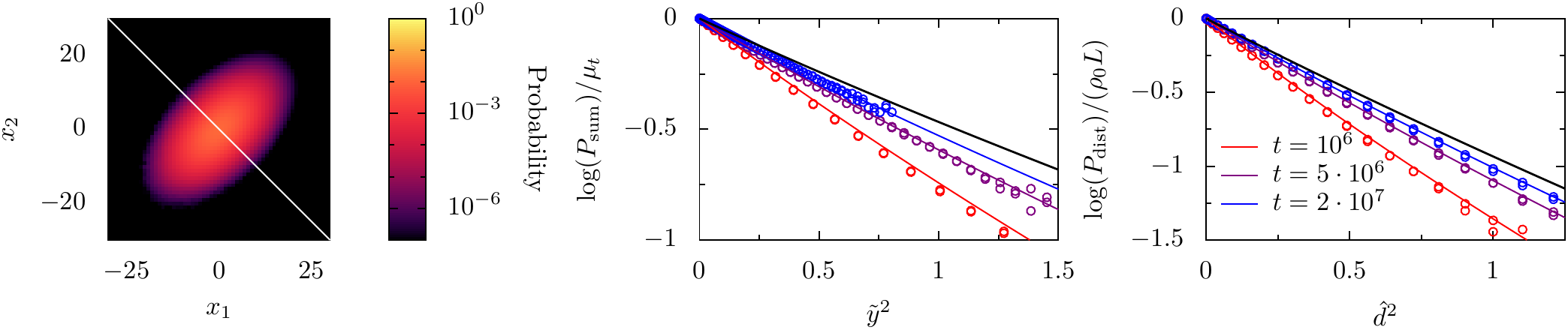}
 \caption{Large deviation functions for two TPs for $\rho=0.01$, $L=10^3$ and $t=10^6$
 {\it Left:} joint probability distribution of $(x_1,x_2)$ at $t=5\cdot 10^6$,
 the bottom-left triangle correponds to the numerical simulations while the top-right triangle
 is the numerical solution of (\ref{eq:ld1}) for two TPs (see SI).
 {\it Center:} rescaled marginal probability density of the half-sum $P_\mathrm{sum}((Y_1+Y_2)/2)$ at times $t=1\cdot 10^6, 5\cdot 10^6, 2\cdot 10^7$
 ($\tau = 1, 5, 20$).
 The dots are the results of numerical simulations
 while the colored lines comes from Eq.~(\ref{eq:ld2TP}).
 The black curve is the prediction when $\tau\to\infty$ (Eq.~(\ref{eq:probCM})).
 {\it Right:} rescaled marginal probability
 density of the distance $P_\mathrm{dist}((X_2-X_1)/2)$ at the same times. The black line comes from Eq.~(\ref{eq:probDist}).
 }
 \label{fig3}
\end{figure*}

{\it Characteristic function in the dense limit.}
Importantly, the characteristic function can be shown from Eq.~\eqref{eq:psiSol} to admit the following simple form in the scaling limit $t\to\infty$, $L=L_1+\dots+L_{N-1}\to\infty$
with fixed rescaled time $\tau = t/L^2$ and fixed relative lengths $\lambda_i^{(1)} = L_i/L$:
\begin{multline} \label{eq:psiResTime}
 \lim_{\rho_0\to 0}\frac{\psi(\vec k, t)}{\rho_0} = \sqrt\frac{2t}{\pi}
\sum_{n=0}^{N-1} \sum_{i=1}^{N-n} (\cos(k_i + \dots + k_{i+n})-1) \times \\
 \bigg[
 g\(\frac{\lambda_i^{(n)}}{\sqrt{2\tau}}\) - g\(\frac{\lambda_{i-1}^{(n+1)}}{\sqrt{2\tau}}\)
 - g\(\frac{\lambda_i^{(n+1)}}{\sqrt{2\tau}}\) + g\(\frac{\lambda_{i-1}^{(n+2)}}{\sqrt{2\tau}}\)
 \bigg]
\end{multline}
with
\begin{equation} \label{eq:funG}
 g(u) = e^{-u^2} - \sqrt\pi\, u\, \text{erfc}(u)
\end{equation}
and  $\lambda_i^{(n)} = (L_i + \dots + L_{i+n-1})/L$ (and $\lambda_i^{(0)} = 0$).
We use the convention $L_0 = L_N = +\infty$ (Fig.~\ref{fig1}).

Equation (\ref{eq:psiResTime})  gives us the full $N$-tag probability law
of the SEP in the dense limit and is the main result of this Letter.
In the following we analyse two important consequences.

{\it Large deviations in the dense limit.}
Noticing that $\psi(-i\vec s, t) = \ln(\mathbb{E}[e^{\vec s\cdot \vec Y(t)}])$, it is possible
to apply the G\"artner-Ellis theorem \cite{Touchette:2009} of large deviations
to get an expression for the joint probability in the large-time limit
$\mu_t = \rho_0\sqrt\frac{2t}{\pi} \to\infty$:
\begin{align} \label{eq:ld1}
 P\(\{\tilde y_i=\mu_t^{-1}Y_i\}\) &\asymp e^{-\mu_t J(\{\tilde y_i\})}
\end{align}
where $J(\vec{\tilde y})$ is the Legendre transform of $\mu_t^{-1} \psi(-i\vec s)$ and 
the symbol '$\asymp$' means equivalence at exponential order.

The probability law is best described using as variables  the ``half-sum of extremal
displacements''
$Y=(Y_1+Y_N)/2$ and the ``distances'' $D_i=(Y_{i+1}-Y_i)/2$.
For two TPs (see \cite{SupplMat} for $N$ TPs),  the large
deviation function is found to be given by: 
\begin{align}\label{eq:ld2TP}
 &J(\tilde y, \tilde d_1) = 
 \sup_{u,v\in\mathbb{R}}
 \bigg\{
 u\tilde y + v\tilde d_1 - \(\cosh(u) - 1\)  \\
 &- \left[1-g\(\frac{1}{\sqrt{2\tau}}\)\right]\(2\cosh\frac{u}{2} \cosh\frac{v}{2} - \cosh(u) - 1\) \bigg\} \nonumber,
\end{align}
where $\tilde y=(\tilde y_1+\tilde y_2)/2$ and $\tilde d_1=(\tilde y_2-\tilde y_1)/2$.
This expression gives the  probability law of the TPs at arbitrary  times.
We checked it against numerical simulations of the random walks of the vacancies
and we found a very good agreement (Fig. \ref{fig3}).

At large rescaled time ($\tau\to\infty$), it is found that 
\begin{align} \label{eq:probCM}
 P\(Y=\mu_t\tilde y, \{D_i=\mu_t\tilde d_i\}\) &\underset{\tau\to\infty}{\asymp} e^{-\mu_t I(\tilde y)}
 \prod_i \delta(\tilde d_i) \\
 P_\text{dist}\(\{D_i=\rho_0 L_i \hat d_i\}\) &\underset{\tau\to\infty}{\asymp} \prod_i e^{-2\rho_0 L_i I(\hat d_i)}
 \label{eq:probDist} \\
 I(u) = 1 - \sqrt{1+u^2} &+ u\ln\left(u + \sqrt{1+u^2}\right). \label{eq:funI}
\end{align}
In particular, we recover that all TPs behave as a single one, as shown above (see Eq.~\eqref{eq:equal1N}).
Indeed, the function $I$ is the large deviation function involved in the dense limit of the single TP problem  as can be extracted from Ref.~\cite{Illien:2013} or from
the limit $\rho\to 1$ of Ref~\cite{Imamura:2017}.
Note also that the marginal law of the distances $P_\text{dist}$ can be
deduced from Eq.~(\ref{eq:lawDist}) when $\rho\to1$ in the particular case of 2 TPs~\cite{SupplMat}.
Finally,  at large times, in the small deviations regime,  
Eq.~(\ref{eq:ld1}) gives back a Gaussian
law with $\langle Y^2\rangle = \rho_0 \sqrt{2t/\pi}$ and $\langle D_i^2\rangle =
2\rho_0 L_i$.

\begin{figure}
 \includegraphics[width=8cm]{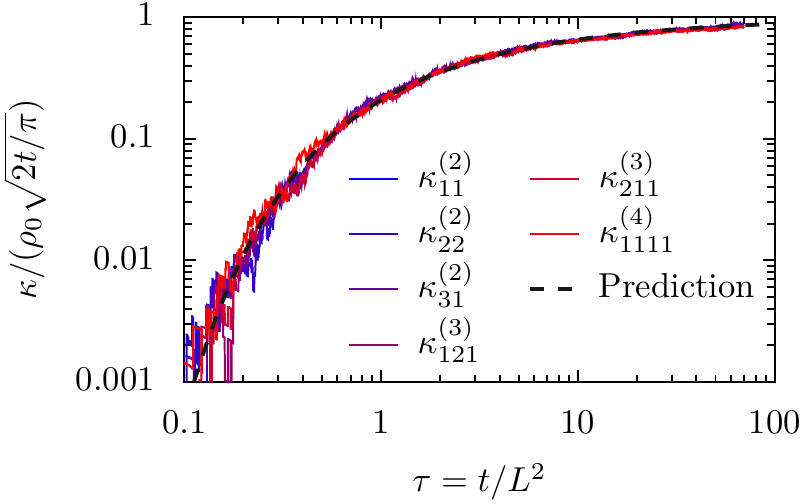}
 \caption{Rescaled evolution of the cumulants associated to two to four TPs.
 $\rho_0 = 0.002$, $L=12$ is the total distance.
 The solid colored curves correspond to the simulations
 in continuous time. The dashed black curve is the prediction Eq.~(\ref{eq:asympCum1}).
 }
 \label{fig2}
\end{figure}

{\it Universal scaling of the cumulants in the dense limit.}
A striking  consequence of Eq.~(\ref{eq:psiResTime}) is that all the even cumulants ($p_1 + \dots + p_N$ even in Eq.~\eqref{eq:defCum}) are equal and assume the universal scaling form (the odd cumulants are equal to zero):
\begin{equation} \label{eq:asympCum1}
 \lim_{\rho_0\to0} \frac{\kappa_{\text{even}}^{(N)}(t)}{\rho_0}
 = \sqrt\frac{2t}{\pi} g\(\frac{1}{\sqrt{2\tau}}\)
 + o(\sqrt{t})
\end{equation}
where $g$ was defined in Eq.~\eqref{eq:funG}. 
Several comments are in order.
(i) This expression is found to be in very good agreement with continuous-time simulations of the SEP at any time (Fig. \ref{fig2}).
(ii)  The leading order in time is the same as the one obtained for a single TP \cite{Illien:2013}, as expected from Eq.~(\ref{eq:equal1N}).
(iii) A major result is that the scaling form is the same
for all cumulants and depends only on the distance $L$ between the first and
the last tagged particles.
(iv) The Edwards-Wilkinson equation, which is seen as the Gaussian limit
of the SEP~\cite{Spohn:1983,Gupta:2007}, provides the first two cumulants and leads to
$\kappa^{(2)}_{11} = \kappa^{(1)}_{2} g\left((2\tau)^{-1/2}\right)$ at any density
\cite{Krapivsky:2009,Majumdar:1991}
(see \cite{SupplMat} for a numerical verification), consistently with Eq.~\eqref{eq:asympCum1}.
(v) A similar scaling for $\kappa^{(2)}_{11}$ has also been found in the
random average process~\cite{Rajesh:2001,Cividini:2016}.
Note that here, this scaling form is shown to hold for {\it all} the cumulants, and for an {\it arbitrary} number of TPs.
 
\textit{Conclusion}
To sum up, we studied the joint probability distribution of an arbitrary number of tagged particles in the SEP. In  the large time limit, we determined  the leading behavior of all cumulants  for an arbitrary density of particles.   We  obtained  the  full dynamics in the dense limit of particles and  explicitly derived  the time dependent large deviation function of the problem. We also unveiled  a universal scaling form shared by all cumulants. We stress that this universal behavior is a non trivial high density effect. The dynamics of the cumulants for an arbitrary density of particles is expected to be non universal and
remains to be determined.


\clearpage

\appendix

\renewcommand{\theequation}{S\arabic{equation}}
\renewcommand{\thefigure}{S\arabic{figure}}

\begin{widetext}
 
\section{Cumulants at arbitrary density}
\subsection{Law of the distance between two particles}
We consider two tagged particles (TPs) in the SEP with density $\rho$. The initial distance
between them is $L$ and we derive the equilibrium distribution of the distance
(in particular we show that the distribution of the distance is time-independant at large time).
We proceed as follow: (i) We write the law of the number of particles $N_p$ between the TPs.
This number is fixed initially and does not evolve. (ii) We write the law of the number
of vacancies $N_v$ between the tracers at equilibrium; this law depends on $N_p$.

Initially there are $L-1$ sites between the TPs. Each site is occupied with probability $\rho$. This gives us a binomial law for $N_p$:
\begin{equation}
 \mathbb{P}(N_p = k) = \binom{L-1}{k} \rho^k (1-\rho)^{L-1-k}
\end{equation}

At large time, the number of vacancies between the two TPs, knowing that there
are $k$ particles between them, is given by
the law of the number $m$ of failures before $k+1$ successes in a game in which the probability of a success is $\rho$.
It is a negative binomial law:
\begin{equation}
 \mathbb{P}(N_v = m|N_p = k) = \binom{m+k}{m} (1-\rho)^m \rho^{k+1}
\end{equation}

The distance $D$ between the tracers is given by $D=N_p + N_l + 1$, its law is:
\begin{align}
 \mathbb{P}(D=\delta) &= \sum_{k=0}^{L-1} \mathbb{P}(N_p = k)\mathbb{P}(N_v = \delta-k-1|N_p = k) \\
 &= \sum_{k=0}^{L-1}\binom{L-1}{k} \rho^k (1-\rho)^{L-1-k}
 \binom{\delta-1}{\delta-k-1} (1-\rho)^{\delta-k-1} \rho^{k+1} \\
 &= \sum_{k=0}^{L-1}\binom{L-1}{k} \binom{\delta-1}{k} \rho^{2k+1} (1-\rho)^{L+\delta-2k-2}
 \label{si:eq:predDist}
\end{align}
This law is in very good agreement with the numerical simulations (Fig.~\ref{si:fig:dist}).

\subsection{Large deviations of the law of the distance}
From the law of the distance \eqref{si:eq:predDist}, one can derive the generating function
$G_D(z)$.
\begin{align}
 G_D(z) &\equiv \sum_{\delta=1}^\infty \mathbb{P}(D=\delta) z^\delta \\
 &= \sum_{\delta=1}^\infty z^\delta \sum_{k=0}^{L-1} \mathbb{P}(N_p = k)\mathbb{P}(N_v = \delta-k-1|N_p = k) \\
 &= z \sum_{k=0}^{L-1} z^k \mathbb{P}(N_p = k) \sum_{m=0}^\infty z^m \mathbb{P}(N_v = m|N_p = k) \\
 &= z \sum_{k=0}^{L-1} z^k \binom{L-1}{k} \rho^k (1-\rho)^{L-1-k}
 \sum_{m=0}^\infty  \binom{m+k}{m} (1-\rho)^m \rho^{k+1} \\
 &= z \sum_{k=0}^{L-1} z^k \binom{L-1}{k} \rho^k (1-\rho)^{L-1-k}
 \left(\frac{\rho}{1-(1-\rho)z}\right)^{k+1} \\
 &= \left(\frac{\rho z}{1-(1-\rho)z}\right) \left(\frac{\rho^2 z}{1-(1-\rho)z} + 1-\rho\right)^{L-1}
\end{align}

We can then derive a large deviation scaling in the limit $L\to\infty$:
\begin{equation}
 \frac{1}{L} \ln G_D(e^t) \xrightarrow[N\to\infty]{} \ln\left(\frac{\rho^2 z}{1-(1-\rho)z} + 1-\rho\right)
 \equiv \phi(t)
\end{equation}
From the Gärtner-Ellis theorem \cite{Touchette:2009}, this implies:
\begin{align}
 P(D=L(1+\tilde d)) &\asymp e^{-LI(\tilde d)} \\
 I(\tilde d) &= \text{sup}_{t\in\mathbb{R}} \(t(1+\tilde d) - \phi(t)\)
 \label{si:eq:SI_extremum}
\end{align}

We now consider the high-density limit: $\rho = 1-\rho_0$ with $\rho_0 \ll 1$. We obtain:
\begin{equation}
 \phi(t) = t + 2\rho_0 \cosh 
\end{equation}
Thus, the supremum in \eqref{si:eq:SI_extremum} is at $t^\ast$ such that:
\begin{equation}
 \sinh t^\ast = \frac{\tilde d}{2\rho_0}
\end{equation}
At the end of the day,
\begin{equation}
 I(\tilde d) = 2\rho_0\left\{
 1-\sqrt{1+\(\frac{\tilde d}{2\rho_0}\)^2}
 + \frac{\tilde d}{2\rho_0} \log\(\frac{\tilde d}{2\rho_0} + \sqrt{1+\(\frac{\tilde d}{2\rho_0}\)^2}\)
 \right\}
\end{equation}
This is exactly what is found with our high-density approach: Eq.~\eqref{eq:probDist}
of the main text.

\subsection{Large time behavior of the cumulants of $N$ particles}
We consider $N$ TPs having displacements $Y_1, \dots Y_N$.
We know that the moments of a single particle scale as $t^{1/2}$ \cite{Imamura:2017}
while the moments of the distance scale as $t^0$ (previous section).
\begin{align}
 \langle Y_1^{2p} \rangle &= \mathcal{O}(t^{1/2})\qquad \forall p \in \mathbb{N} \label{si:eq:SI_X1} \\
 \langle (Y_i - Y_1)^{2p} \rangle &= \mathcal{O}(t^0)\qquad \forall i \leq N, \forall p \in \mathbb{N}
 \label{si:eq:SI_dist}
\end{align}

From this we want to show that
\begin{equation} \label{si:eq:SI_HR}
 A_{p_1, \dots p_N}^{(N)} \equiv \langle Y_1^{p_1}\dots Y_N^{p_N} \rangle 
 - \langle Y_1^{p_1 + \dots + p_N} \rangle = \mathcal{O}(t^{1/4}) \qquad
 \forall p_1, \dots, p_N
\end{equation}

We proceed by induction: the case $N=1$ is straightforward. Now assuming that
(\ref{si:eq:SI_HR}) holds for a given $N$, we want to prove it for $N+1$.
As we have $A_{p_1, \dots p_N, 0}^{(N+1)} = A_{p_1, \dots p_N}^{(N)} = \mathcal{O}(t^{1/4})$,
we prove by induction that
$A_{p_1, \dots p_N, q}^{(N+1)}= \mathcal{O}(t^{1/4})\ \forall q\geq 0$.
Indeed, if $A_{p_1, \dots p_N, q'}^{(N+1)}= \mathcal{O}(t^{1/4})\ \forall q'<q$, we can
write:
\begin{align}
 &A_{p_1, \dots p_N, q}^{(N+1)} =
 \langle Y_1^{p_1}\dots Y_N^{p_N}X_{N+1}^q \rangle - \langle Y_1^{p_1 + \dots + p_N + q} \rangle \\
  &= \left\langle Y_1^{p_1}\dots Y_N^{p_N}\left[(Y_{N+1}-Y_1)^q
  -\sum_{r=1}^q \binom{q}{r} (-1)^r Y_1^r Y_{N+1}^{q-r}\right] \right\rangle
  - \langle Y_1^{p_1 + \dots + p_N + q} \rangle \\
  &= \langle Y_1^{p_1}\dots Y_N^{p_N}(Y_{N+1}-Y_1)^q \rangle
  -\sum_{r=1}^q \binom{q}{r} (-1)^r \langle Y_1^{p_1+r} Y_2^{p_2} \dots Y_{N+1}^{q-r}\rangle 
  - \langle Y_1^{p_1 + \dots + p_N + q} \rangle \\
  &= \langle Y_1^{p_1}\dots Y_N^{p_N}(Y_{N+1}-X_1)^q \rangle
  -\sum_{r=1}^q \binom{q}{r} (-1)^r \left[ \langle Y_1^{p_1+r} Y_2^{p_2} \dots Y_{N+1}^{q-r}\rangle
  - \langle Y_1^{p_1 + \dots + p_N + q}\rangle \right]
\end{align}
All the terms in the sum are of order $\mathcal{O}(t^{1/4})$ from (\ref{si:eq:SI_HR})
and the first term can be bounded by the Cauchy–Schwarz inequality:
\begin{equation}
 \left|\langle Y_1^{p_1}\dots Y_N^{p_N}(Y_{N+1}-Y_1)^q \rangle\right| \leq
 \sqrt{\langle \(Y_1^{p_1}\dots Y_N^{p_N}\)^2 \rangle\langle (Y_{N+1}-Y_1)^{2q} \rangle}
 = \sqrt{\mathcal{O}(t^{1/2})\mathcal{O}(t^0)} = \mathcal{O}(t^{1/4})
\end{equation}
using (\ref{si:eq:SI_X1}), (\ref{si:eq:SI_dist}) and (\ref{si:eq:SI_HR}).
This ends the proof of (\ref{si:eq:SI_HR}).

This implies that if $p_1 + \dots + p_N$ is even:
\begin{equation}
  \langle Y_1^{p_1}\dots Y_N^{p_N} \rangle \equi{t\to\infty} \langle Y_1^{p_1 + \dots + p_N}
  \rangle = \mathcal{O}(t^{1/2})
\end{equation}
The moments of $N$ particles are equal, in the large time limit, are given by
the moments of a single particle.
This extends to the cumulants $\kappa^{(N)}_{p_1, \dots, p_N}$ defined from the second
characteristic function.
\begin{equation}
 \psi(k_1, \dots k_N) \equiv \sum_{p_1, \dots p_N} \kappa^{(N)}_{p_1, \dots, p_N}
  \frac{(ik_1)^{p_1} \dots (ik_N)^{p_N}}{p_1!\dots p_N!}
  \equiv \log\left[\sum_{p_1, \dots p_N} \langle Y_1^{p_1}\dots Y_N^{p_N} \rangle
  \frac{(ik_1)^{p_1} \dots (ik_N)^{p_N}}{p_1!\dots p_N!}\right]
\end{equation}
\begin{equation}
  \kappa^{(N)}_{p_1, \dots, p_N} \equi{t\to\infty} \kappa^{(1)}_{p_1 + \dots + p_N}
  \equi{t\to\infty} B_{p_1+\dots + p_N} \sqrt{t}
\end{equation}
The coefficient $B_p$ are computed in Ref. \cite{Imamura:2017}.

\begin{figure}
 \includegraphics[width=15cm]{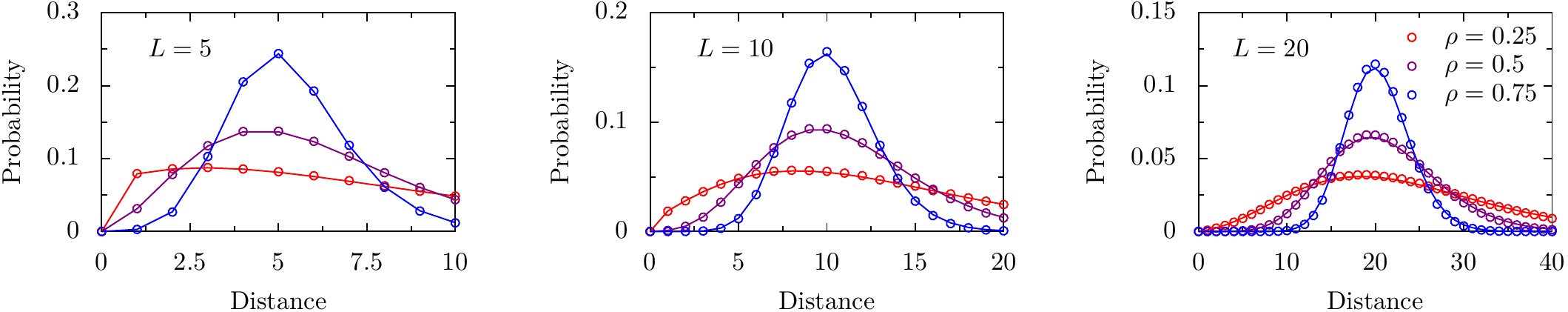}
 \caption{Comparison of the numerical probability distribution of the distance of two TPs
 (dots)
 with the prediction \eqref{si:eq:predDist} (lines) for $\rho = 0.25, 0.5, 0.75$ and
 $L=5, 10, 20$. The average is performed other $10^6$ simulations at final time
 $t=2\cdot 10^4$ (we checked that this is enough for the convergence).}
 \label{si:fig:dist}
\end{figure}

\section{Detailed calculations in the high-density limit}
\subsection{Approximation and thermodynamic limit}
Let us consider a system of size $\mathcal{N}$ with $M$ vacancies
and denote by $\vec Y(t) = (X_i(t)-X_i^0)_{i=1}^N$
the vector of the displacements of the TPs.
The probability $P^{(t)} (\vec Y|\{Z_j\})$ of having displacements $\vec Y$ at
time $t$ knowning that the $M$ vacancies started at sites $Z_1\dots Z_M$ is exactly given by~:
\begin{equation}
 P^{(t)} (\vec Y|\{Z_j\}) =
  \sum_{\vec Y_1,\dots,\vec Y_M} \delta_{\vec Y, \vec Y_1+\dots+\vec Y_M}
  \mathcal{P}^{(t)} (\{\vec Y_j\}|\{Z_j\})
\end{equation}
where $\mathcal{P}^{(t)} (\{\vec Y_j\}|\{Z_j\})$ is the probability of displacement $\vec Y_j$ due to the vacancy $j$
for all $j$ knowing the initial positions of all the vacancies.

Assuming that in the large density large ($\rho\to 1$) the vacancies interact
independantly with the TPs, we can 
link it to the probability $p^{(t)}_Z(\vec Y)$ that the tracers have moved by $\vec Y$ at time $t$
due to a single vacancy that was initially at site $Z$:
\begin{equation}
 \mathcal{P}^{(t)} (\{\vec Y_j\}|\{Z_j\}) \underset{\rho\to 1}{\sim} 
  \prod_{j=1}^M p^{(t)}_{Z_j}(\vec Y_j)
\end{equation}
so that
\begin{equation}\label{si:eq:approx}
 P^{(t)} (\vec Y|\{Z_j\}) \underset{\rho\to 1}{\sim}
  \sum_{\vec Y_1,\dots,\vec Y_M}
  \delta_{\vec Y, \vec Y_1+\dots+\vec Y_M}
  \prod_{j=1}^M p^{(t)}_{Z_j}(\vec Y_j)
\end{equation}

We take the Fourier transform and we average over the initial positions of the vacancies:
\begin{equation}
\tilde p^{(t)}(\vec k) \equiv \frac{1}{\mathcal{N}-N}
\sum_{Z \notin \{X_i^0\}} \sum_{\vec Y} p^{(t)}_Z(\vec Y) e^{i\vec k\vec Y}
\end{equation}
and mutatis mutandis for $\tilde P^{(t)} (\vec k)$.
Furthermore we write $\tilde p_Z^{(t)}(\vec k) = 1 + \tilde q_Z^{(t)}(\vec k)$ ($q_Z$ corresponds to the deviation
from a Dirac centered in $0$).
\begin{equation}
 \tilde P^{(t)} (\vec k) \underset{\rho\to 1}{\sim} \left[\tilde p^{(t)}(\vec k)\right]^M
 = \left[\frac{1}{\mathcal{N}-N}\sum_{Z\notin \{X_i^0\}}\tilde p_Z^{(t)}(\vec k)\right]^M
 = \left[1 + \frac{1}{\mathcal{N}-N}\sum_{Z\notin \{X_i^0\}} \tilde q_Z^{(t)}(\vec k)\right]^M
\end{equation}

We now take the limit $\mathcal{N},M\to\infty$
with $\rho_0 \equiv 1-\rho = M / \mathcal{N}$ (density of vacancies) remaining constant.
The second characteristic function reads:

\begin{equation} \label{si:eq:psi1}
 \lim_{\rho_0\to 0} \frac{\psi^{(t)}(\vec k)}{\rho_0} \equiv
 \lim_{\rho_0\to 0} \frac{\ln \left[\tilde P^{(t)}(\vec k)\right]}{\rho_0}
 = \sum_{Z \notin \{X_i^0\}} \tilde q_Z^{(t)}(\vec k)
\end{equation}

\subsection{Expression of the single-vacancy propagator}
One can partition over the first passage of the vacancy
to the site of one of the tracers to get an expression for $\tilde q_Z$ which is the main quantity involved
in (\ref{si:eq:psi1}):
\begin{align}
 p_Z^{(t)}(\vec Y) &= \delta_{\vec Y, \vec 0}  \(1 - \sum_{j=0}^t \sum_{\nu = \pm 1, \pm 2} F^{(j)}_{\nu, Z}\)
 + \sum_{j=0}^t \sum_{\nu = \pm 1, \pm 2} p_{-\nu}^{(t-j)}(\vec Y) F^{(j)}_{\nu, Z} \\
 \tilde p_Z^{(t)}(\vec k) &= 1 - \sum_{j=0}^t \sum_{\nu = \pm 1, \pm 2}
 F^{(j)}_{\nu, Z} + \sum_{j=0}^t \sum_{\nu = \pm 1, \pm 2}
 \tilde p_{-\nu}^{(t-j),\zeta} (\vec k) F^{(j)}_{\nu, Z} \\
\label{si:eq:qAll}
 \tilde q_Z^{(t)}(\vec k) &= - \sum_{j=0}^t \sum_{\nu = \pm 1, \pm 2}
 \left[1 - \(1+\tilde q_{-\nu}^{(t-j), \zeta} (\vec k)\) F^{(j)}_{\nu, Z}\right]
\end{align}
An exponant $\zeta$ to a quantity means that this quantity
is computed taking into account $\zeta = \zeta(Z)$.

We now need an expression for $\tilde q_\eta^\zeta$ where $\eta$ is a special site.
To do so we decompose the propagator of the displacements over the successive passages of the vacancy to
the position of one of the tracers:
\begin{multline} \label{si:eq:decompo}
 p^{(t),\zeta}_\eta (\vec Y) =
 \delta_{\vec Y,\vec 0} \(1 - \sum_{j=0}^t \sum_\mu F^{(j),\zeta}_{\mu, \eta}\) \\
  + \sum_{p=1}^\infty \sum_{m_1,\dots,m_p=1}^{\infty} \sum_{m_{p+1}=0}^\infty
 \delta_{t, \sum_i\! m_i} \sum_{\nu_1, \dots, \nu_p}
  \delta_{\vec Y, \sum_i\! \vec e_{\nu_i}}
   \(1- \sum_{j=0}^{m_{p+1}} \sum_\mu F^{(j),\zeta}_{\mu, -\nu_p}\)
  F^{(m_p),\zeta}_{\nu_p, -\nu_{p-1}} \dots F^{(m_2),\zeta}_{\nu_2, -\nu_1}
  F^{(m_1),\zeta}_{\nu_1, \eta}
\end{multline}
the sums on $\mu$ and $\nu_i$ run over the special sites ($\pm 1, \dots \pm N$).

The discrete Laplace transform (power series) of a function of time $g(t)$ is
$\hat g(\xi)\equiv \sum_{t=0}^\infty g(t) \xi^t$.
We can now take both the Laplace and Fourier transforms of (\ref{si:eq:decompo}) to get:
\begin{equation}
 \hat p_{\eta}^\zeta(\vec Y, \xi) = \frac{1}{1-\xi}\left\{ \delta_{\vec Y,\vec 0}
 \(1 - \sum_\mu \hat F_{\mu,\nu}^\zeta\)
 + \sum_{p=1}^\infty \sum_{\nu_1, \dots, \nu_p} \delta_{\vec Y, \sum_i\! \vec e_{\nu_i}}
 \sum_\mu \(1-\hat F_{\mu,-\nu_p}^\zeta\) \hat F_{\nu_p, -\nu_{p-1}}^\zeta
 \dots \hat F_{\nu_2, -\nu_1}^\zeta \hat  F_{\nu_1, \eta}^\zeta 
 \right\}
\end{equation}

\begin{equation} \label{si:eq:qBorder}
 \hat{\tilde q}_\eta^\zeta (\vec k, \xi)
 \equiv \hat{\tilde p}_\eta^\zeta(\vec k, \xi) - \frac{1}{1-\xi}
 = \frac{1}{1-\xi} \sum_{\mu,\nu}
 \{[1-T^\zeta(\vec k, \xi)]^{-1}\}_{\nu\mu} 
 \times \(1-e^{-i\,\vec k\vec e_\nu}\) e^{i\,\vec k\vec e_\mu}
 \hat F^\zeta_{\mu\eta}(\xi)
\end{equation}

The matrix $T$ is defined by
$T^\zeta(\vec k,\xi)_{\nu\mu} = \hat F^\zeta_{\nu,-\mu} (\xi) e^{i\,\vec k \vec e_{\nu}}$.

\subsection{Expression of the characteristic function}
Introducing (\ref{si:eq:qAll}) into (\ref{si:eq:psi1}) directly gives:
\begin{align}
 \lim_{\rho_0\to 0}\frac{\hat \psi (\vec k, \xi)}{\rho_0} &= - \sum_\nu \bigg[
 \frac{1}{1-\xi}
 - \(\frac{1}{1-\xi} + \hat{\tilde q}^{(\zeta)}_{-\nu} (\vec k, \xi)\)
 e^{i\,\vec k\vec e_\nu} \bigg] h_\zeta(\xi) \\
 h_\zeta(\xi) &= \sum_{Z\notin \{X_i^0\}} \hat F_{\zeta,Z} (\xi)
 = \sum_{Z=X_\zeta^0 + 1}^{X_{\zeta+1}^0 - 1} \hat F_{\zeta,Z} (\xi)
\end{align}
with $\hat{\tilde q}_{\nu}^\zeta$ given by (\ref{si:eq:qBorder}).

\subsection{Expression of the quantities of interest
($\hat F_{\nu, \eta}^{(\zeta)}$ and $h_\mu$)}
The two results \cite{Hughes:1995} that we recalled in the article are:
\begin{itemize}
 \item The Laplace transform of the first passage density at the origin at time $t$
 of a symmetric 1d Polya walk starting from site $l$ is:
 \begin{equation} \label{si:eq:walk1}
  \hat f_l(\xi) = \alpha^{|l|} \mbox{ with } \alpha = \frac{1-\sqrt{1-\xi^2}}{\xi}
 \end{equation}
  \item The probability of first passage at site $s_1$ starting from $s_0$
  and considering $s_2$ as an absorbing site is given by
\begin{equation} \label{si:eq:walk2}
 \hat F^\dagger (s_1|s_0,\xi) =
 \frac{\hat f_{s_1-s_0}(\xi) - \hat f_{s_1-s_2}(\xi) \hat f_{s_2-s_0}(\xi)}
 {1-\hat f_{s_1-s_2}(\xi)^2}
\end{equation}
\end{itemize}

From (\ref{si:eq:walk1}), we have:
\begin{equation}
\hat F_{-1,-1} = \hat F_{+N, +N} = \alpha
\end{equation}

From (\ref{si:eq:walk1}, \ref{si:eq:walk2}), and recalling that the distance between
TP $\mu$ and TP $\mu+1$ is $L_\mu^{(\zeta)}$, we have
\begin{align}
 \hat F_{\mu,\mu}^{(\zeta)} = \hat F_{-\mu+1,-\mu-1}^{(\zeta)} &=
 \frac{\alpha -\alpha^{2L_\mu^{(\zeta)}-1}}{1-\alpha^{2L_\mu^{(\zeta)}}} \\
 \hat F_{\mu,-\mu+1}^{(\zeta)} = \hat F_{-\mu+1,\mu}^{(\zeta)} &= 
 \frac{\alpha^{L_\mu^{(\zeta)}-1} -\alpha^{L_\mu^{(\zeta)} + 1}}{1-\alpha^{2L_\mu^{(\zeta)}}} 
\end{align}

And the sums are:
\begin{align}
 h_0 = h_N &= \sum_{Z=-\infty}^{-1} \alpha^{|Z|} = \frac{\alpha}{1-\alpha} \\
 h_\mu &= \sum_{Z=1}^{L_\mu-1} \frac{\alpha-\alpha^{2L_\mu-Z}}{1-\alpha^{2L_\mu}}
 = \frac{\alpha (1-\alpha^{L_\mu-1})(1-\alpha^{L_\mu})}{(1-\alpha)(1-\alpha^{2L_\mu})}
 \label{si:eq:exprEnd} 
\end{align}

\section{Results in the high density limit}
\subsection{Scaling of the characteristic function}
Using a numerical software (Mathematica)
we obtain the following result for the discrete Laplace tranform
of the second characteristic function is:
\begin{multline} \label{si:eq:psiResSI}
 \lim_{\rho_0\to 0}\frac{\hat \psi(\vec k, \xi)}{\rho_0} = \frac{1}{(1-\alpha^2)(1-\xi)}
\sum_{n=0}^{N-1} \sum_{i=1}^{N-n}
 \alpha^{\mathcal{L}_i^n}
 \bigg\{ 
 2\alpha (1-\alpha^{L_{i-1}})(1-\alpha^{L_{i+n}}) \cos(k_i + \dots + k_{i+n}) \\
 + (1-\alpha) Q_n(k_i, \dots, k_{i+n}) + C
 \bigg\}
\end{multline}
with $C$ a constant enforcing $\hat\psi(\vec k =\vec 0) = 0$ and
\begin{align}
 \mathcal{L}_i^n &= L_i + \dots + L_{i+n-1} \\
 Q_2(k_1, k_2) &= \alpha^{L_1}\(e^{ik_1} + e^{-ik_2}\) \label{si:eq:SI_Q2} \\
 Q_3(k_1, k_2, k_3) &= \alpha^{L_1}\(e^{ik_1} + e^{-ik_2}\)
 + \alpha^{L_2}\(e^{ik_2} + e^{-ik_3}\)
 + \alpha^{L_1+L_2}\(e^{i(k_1+k_2)} + e^{-i(k_2+k_3)} + 2\cos k_2\)
\end{align}
Similar expressions exist for $Q_n, n>3$.

We define the total length $L = L_1 + \dots + L_{N-1}$ and the rescaled variables
$l_i^n = \mathcal{L}_i^n/L$.

We define $p=1-\xi$ and $\tilde p = p L^2$. We take the limit
$L\to\infty$ keeping $\tilde p$ constant. This is a limit of large time.

The asymptotic behavior of $\alpha$ is given by~:
\begin{equation}
 \alpha = \frac{1-\sqrt{1-\xi^2}}{\xi} 
 = 1-\sqrt{2p} + \mathcal{O}_{p\to 0}(p)
\end{equation}
so that
\begin{equation}
 \alpha^{rL} \underset{L\to\infty}{\sim} e^{-r\sqrt{2\tilde p}}
\end{equation}

(\ref{si:eq:psiResSI}) becomes~:
\begin{multline} \label{si:eq:psiResSI2}
 \lim_{\rho_0\to 0}\frac{\hat \psi(\vec k, p=L^2\tilde p)}{\rho_0}
 \underset{L\to\infty}{\sim}
 \frac{L^3}{\sqrt{2} \tilde p^{3/2}} \sum_{n=0}^{N-1} \sum_{i=1}^{N-n}
 \(e^{-\sqrt{2\tilde p}\, l_i^n} - e^{-\sqrt{2\tilde p}\, l_{i-1}^{n+1}}
 - e^{-\sqrt{2\tilde p}\, l_i^{n+1}} + e^{-\sqrt{2\tilde p}\, l_{i-1}^{n+2}}
 \) \\
 \times \(\cos(k_i + \dots + k_{i+n}) - 1\)
\end{multline}

The following continuous inverse Laplace tranform is known:
\begin{equation}
 \hat h(p) = \frac{e^{-r\sqrt{2p}}}{p^{3/2}} \Leftrightarrow
 h(t) = 2\sqrt\frac{t}{\pi} g\(\frac{r}{\sqrt{2t}}\)
\end{equation}
\begin{equation}
 g(u) = e^{-u^2} - \sqrt\pi\, u\, \erfc(u)
\end{equation}

(\ref{si:eq:psiResSI2}) can then be inverted and we find:
\begin{multline}
 \lim_{\rho_0\to 0}\frac{\psi(\vec k, t)}{\rho_0}
 \underset{t\to\infty}{\sim} \sqrt\frac{2t}{\pi}
\sum_{n=0}^{N-1} \sum_{i=1}^{N-n}
 \left[
 g\(\frac{L_i^n}{\sqrt{2t}}\) - g\(\frac{L_{i-1}^{n+1}}{\sqrt{2t}}\)
 - g\(\frac{L_i^{n+1}}{\sqrt{2t}}\) + g\(\frac{L_{i-1}^{n+2}}{\sqrt{2t}}\)
 \right] \\
 \times \(\cos(k_i + \dots + k_{i+n}) - 1\)
\end{multline}

\subsection{Effects of the initial conditions on the odd cumulants}
We consider two TPs at an initial distance $L$. We show that, due to the fact that
we impose their initial positions, the two TPs separate slightly from one another.

From (\ref{si:eq:psiResSI}) and (\ref{si:eq:SI_Q2}) we obtain:
\begin{equation}
 \lim_{\rho_0\to 0} \frac{\langle X_2\rangle (\xi)}{\rho_0}
 = \frac{\alpha^L}{(1+\alpha)(1-\xi)} \equi{L\to\infty} L^2
 \frac{e^{-\sqrt{2\tilde p}}}{2\tilde p}
\end{equation}
with $\tilde p = (1-\xi)/L^2$.

The inversion of the Laplace transform gives:
\begin{equation}
 \lim_{\rho_0\to 0} \frac{\langle X_2\rangle (t)}{\rho_0}
 \equi{t\to\infty} \frac{1}{2} \text{erfc}\(\frac{L}{\sqrt{2t}}\)
\end{equation}
and similarly
\begin{equation}
 \lim_{\rho_0\to 0} \frac{\langle X_1\rangle (t)}{\rho_0}
 \equi{t\to\infty} -\frac{1}{2} \text{erfc}\(\frac{L}{\sqrt{2t}}\)
\end{equation}
We indeed see that the two tracers separate a little bit. This effect is obvious
when $L=1$: initially the two TPs are on neighboring sites, and at large time
there is a probability $\rho_0$ that there is a vacancy inbetween: $\langle X_2-X_2\rangle(t\to\infty) = \rho_0$.

Similar effects are expected on all the odd cumulants in the $N$-tag problem.
These effect also add a term of order $t^0$ in the even cumulants.

\subsection{Large deviation function for a single TP}
For a single tracer at high density our approach (which coincides with \cite{Illien:2013}) gives:
\begin{equation}
 \psi(k, t) = \mu_t \(\cos k - 1\)
\end{equation}
with $\mu_t = \rho_0 \sqrt\frac{2t}{\pi}$.
The G\"artner-Ellis theorem gives:
\begin{align} \label{si:eq:SM_ld1}
 P\(Y_i=\mu_t\tilde y\) &\asymp e^{-\mu_t I(\tilde y)} \\
 I(\tilde y) &= \sup_{q\in\mathbb{R}}
 \(q \tilde y - (\cosh q - 1)\)
\end{align}

Solving for the extremum gives:
\begin{equation}
 e^{\pm q} = \pm \tilde y + \sqrt{1 + \tilde y^2}
\end{equation}
and finally:
\begin{equation}
I(\tilde y) = 1 - \sqrt{1+\tilde y^2}
 + \tilde y\ln\left[\tilde y + \sqrt{1+\tilde y^2}\right] 
\end{equation}

\subsection{Large deviation function for $N$ TPs}
We assume that the limits $\rho_0\to 0$ and $t\to\infty$ can be exchanged
(this would need to be proven) and we write
\begin{multline} \label{si:eq:psiResSI3}
 \psi(i\vec q, t)
 \underset{\substack{\rho_0\to 0\\ t\to\infty}}{\sim} \mu_t
\sum_{n=0}^{N-1} \sum_{i=1}^{N-n}
 \left[
 g\(\frac{\Lambda_i^n}{\sqrt\pi}\) - g\(\frac{\Lambda_{i-1}^{n+1}}{\sqrt\pi}\)
 - g\(\frac{\Lambda_i^{n+1}}{\sqrt\pi}\) + g\(\frac{\Lambda_{i-1}^{n+2}}{\sqrt\pi}\)
 \right] \\
 \times \(\cosh(q_i + \dots + q_{i+n}) - 1\)
\end{multline}
with $\mu_t = \rho_0\sqrt\frac{2t}{\pi}$ and $\Lambda_i^n = L_i^n \sqrt\frac{\pi}{2t}$.
In the following, the limits will be implicit.

The Gärtner-Ellis theorem \cite{Touchette:2009} for $\mu_t\to\infty$
then gives us the probability distribution (at exponential order, denoted $\asymp$).
\begin{align}
 P\(\{Y_i = \mu_t\tilde y_i\}\) &\asymp e^{-\mu_t J(\{\tilde y_i\})} \\
 J(\{\tilde y_i\}) &= \sup_{\{q_i\}\in\mathbb{R}}
 \(\sum_{i=1}^N q_i \tilde y_i - \mu_t^{-1} \psi(-i\vec q)\) \label{si:eq:funJSI1}
\end{align}

To simplify the problem we define the following variables:
\begin{equation}
 Y = \frac{Y_1 + Y_N}{2} \qquad D_i = \frac{Y_{i+1}-Y_i}{2}\quad (i=1,\dots N-1)
\end{equation}
\begin{equation}
 u = q_1+\dots+q_N \qquad v_i = -q_1 - \dots - q_i + q_{i+1} + \dots + q_N\quad (i=1,\dots N-1)
\end{equation}

\textbf{Small rescaled length.}
\begin{equation}
 g(\Lambda/\sqrt{\pi}) = 1 - \Lambda + \mathcal{O}_{\Lambda\to 0}(\Lambda^2)
\end{equation}
At the first order in the rescaled lengths, only the ``boundary'' terms contribute in
(\ref{si:eq:psiResSI3}). Using the variables defined above, one checks that (\ref{si:eq:funJSI1})
becomes simpler:
\begin{multline}
 J(\tilde y, \{\tilde d_i\})
 \underset{\Lambda_i \ll 1}{\sim}
 \sup_{u,\{v_i\}\in\mathbb{R}}
 \bigg\{
 u\tilde y + \sum_{i=1}^{N-1} v_i\tilde d_i - \(\cosh(u) - 1\) 
 - \sum_{i=1}^{N-1} \Lambda_i\(2\cosh\frac{u}{2} \cosh\frac{v_i}{2} - \cosh(u) - 1\) \bigg\}
\end{multline}

This extremum can only be solved for in the special cases defined in the main text:
$\Lambda_i = 0$, marginal probability of the distances (this corresponds to $u=0$)
and Gaussian limit ($\tilde y \ll 1$, $\tilde v_i \ll 1$).

\textbf{Two tracers, arbitrary length.}
One should note that for two tracers, (\ref{si:eq:funJSI1}) is already rather simple
if written with the right variables: we don't need to assume $\Lambda\ll 1$.
\begin{equation}
 J(\tilde y, \tilde d) = 
 \sup_{u,v\in\mathbb{R}}
 \bigg\{
 u\tilde y + v\tilde d - \(\cosh(u) - 1\) 
 - \(1-g\(\frac{\Lambda}{\sqrt{\pi}}\)\)\(2\cosh\frac{u}{2} \cosh\frac{v}{2} - \cosh(u) - 1\) \bigg\}
\end{equation}
This expression is used to provide a numerical predition in the main text.

\subsection{Numerical verification of the expression of $\kappa^{(2)}_{11}$ at arbitrary
density}

From the Edwards-Wilkinson equation, one expects~:
\begin{equation}\label{si:eq:cumEW}
 \kappa^{(2)}_{11} = \kappa^{(1)}_{2} g\(\frac{1}{\sqrt{2\tau}}\)
 = \frac{1-\rho}{\rho} \sqrt\frac{2t}{\pi} g\(\frac{1}{\sqrt{2\tau}}\)
\end{equation}
This behavior is in good agreement with numerical simulations (see Fig.~\ref{si:fig:cumInterm}, left).

For the other cumulants, we showed that
\begin{equation} \label{si:eq:equal1N}
\lim_{t\to\infty}\frac{\kappa^{(N)}_{p_1, \dots, p_N}}{\sqrt{t}}
= \lim_{t\to\infty}\frac{\kappa^{(1)}_{p_1 + \dots + p_N}}{\sqrt{t}}
= B_{p_1+\dots+p_N},
\end{equation}
where the constants $B_{k}$ characteristic of a single tracer have been determined in Ref.
\cite{Imamura:2017}.

\begin{figure}
 \includegraphics[width=15cm]{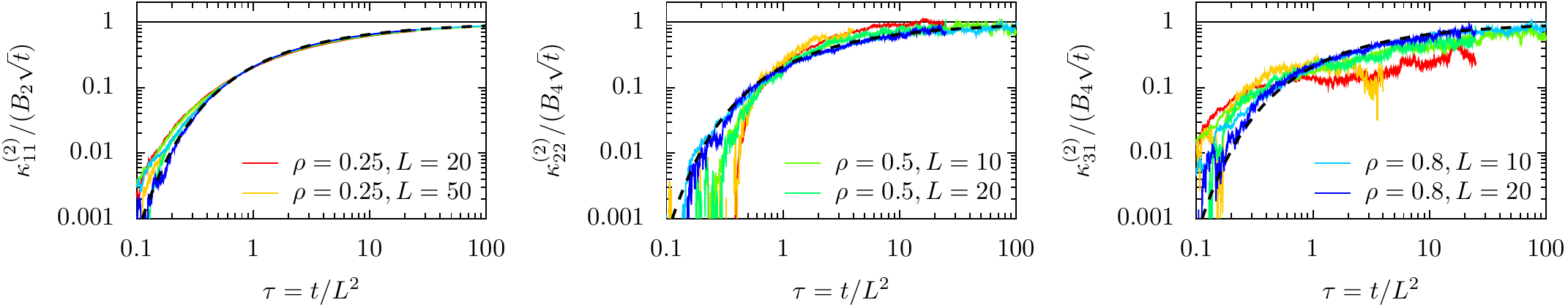}
 \caption{Evolution of the cumulants associated to two TPs
 at different densities and different distances. The cumulants are
 rescaled by the single-tag cumulants, following \eqref{si:eq:cumEW}.
 The dashed line corresponds to $g((2\tau)^{-1/2})$. }
 \label{si:fig:cumInterm}
\end{figure}

\subsection{Argument for the breaking of our scaling shape at arbitrary density}
In analogy with \eqref{si:eq:cumEW}, one would like to be able to make the following
bold conjecture~:
\[\kappa^{(2)}_{p, 2n-p} = B_{2n} \sqrt{t}\, g\((2\tau)^{-1/2}\right)\]
Unfortunately we show that this is incompatible with the law of the distance \eqref{si:eq:predDist}.

Let us focus on the 4th cumulant of the distance (we denote $\langle\rangle_c$ the cumulants)~:
\begin{equation}
  \langle (X_2-X_1)^4 \rangle_c = \langle X_1^4\rangle_c + \langle X_2^4\rangle_c
  -4\langle X_1^3 X_2 \rangle_c - 4\langle X_1 X_2^3 \rangle_c + 6\langle X_1^2 X_2^2 \rangle_c
\end{equation}
We \textit{assume} that~:
\begin{align}
 \langle X_1^4\rangle_c = \langle X_2^4\rangle_c &= B_4\sqrt{t} + o(1) \\
 \langle X_1^3 X_2\rangle_c = \langle X_1 X_2^3\rangle_c = \langle X_1^2 X_2^2\rangle_c
 &= B_4\sqrt{t} g\(\frac{1}{\sqrt{2\tau}}\) = B_4\sqrt{t} - B_4\sqrt\frac{\pi}{2} L + o(1)
\end{align}
This leads us to:
\begin{equation} \label{si:eq:assumeEq}
  \langle (X_2-X_1)^4 \rangle_c = \sqrt{2\pi} B_4 L
\end{equation}

From Ref.~\cite{Imamura:2017},
\begin{equation}
B_4 = \sqrt\frac{2}{\pi} \frac{1-\rho}{\rho^3}
\left[1 - (4-(8-3\sqrt{2})\rho)(1-\rho) + \frac{12}{\pi} (1-\rho)^2\right]
\end{equation}
while from \eqref{si:eq:predDist},
\begin{equation}
  \langle (X_2-X_1)^4 \rangle_c \equi{L\to\infty} 2L \frac{1-\rho}{\rho^3}
  \(12 - 24\rho + 13\rho^2\)
\end{equation}
At an arbitrary density, this is inconsistent with \eqref{si:eq:assumeEq},
thus our conjecture must be wrong. Note that \eqref{si:eq:assumeEq} does hold as expected when $\rho\to 1$.

In Fig.~\ref{si:fig:cumInterm} center (resp. right), we tried the following guess: 
$\kappa^{2}_{22} = B_4\sqrt{t}\, g\((2\tau)^{-1/2}\right)$ (resp. 
$\kappa^{2}_{31} = B_4\sqrt{t}\, g\((2\tau)^{-1/2}\right)$). We see that
it is valid only at high density, as expected.

\section{Description of numerical simulations}
\subsection{Continuous time simulations on a lattice (for the cumulants)}
$N_\text{parts}$ particles are put uniformly at random on the line of size $N_\text{size}$,
except the $N$ tagged particles which are put deterministically on their initial positions.
We used $N_\text{size}=5000$

Each particle has an exponential clock of time constant $\tau = 1$. Thus, the whole
system has an exponential clock of time constant $\tau_\text{all} = \tau / N_\text{parts}$.
When it ticks, a particle is chosen at random and tries to move either to the left or
to the right with probability $1/2$. If the arrival site is already occupied the
particle stays where it was.

The cumulants of the $N$ TPs are averaged over 100\,000 to 500\,000 simulations to obtain
their time dependence.

\subsection{Vacancy-based simulations (for the probability distribution)}
The previous approach does not enable one to get sufficient statistics
to investigate the probability law.

In the case of ponctual brownian particles, Ref.~\cite{Sabhapandit:2015} was able to
use the propagator of the displacement to directly obtain the state of the system at
a given time and investigate the probability distribution.

Here we used a numerical scheme close to out theoretical approach:
at high density and in discrete time, we simulate the behavior of the vacancies
considered as independant random walker. The displacement $\Delta x$ of a vacancy
at (discrete) time $t$ is given by a binomial law:
\begin{align}
 \Delta x &= 2 n_\text{right} - t \\
 P(n_\text{right}, t) &= \frac{1}{2^t} \binom{t}{n_\text{right}}
\end{align}
One is able to recover the final positions of the TPs from the final positions of
the vacancies.

For two TPs at distance $L$, we put a vacancy at each site between the TPs with
probability $\rho_0$ (density of vacancies). We consider a number of sites
$N_\text{sites}$ ($N_\text{sites} = 100\, 000$) on the left of the first TP and
on the right of the second TP and we put a deterministic number of vacancies
$N_\text{vac} = \rho_0 N_\text{sites}$ at random positions on these sites.

We make $10^8$ repetition of the simulation before outputting the probability law.

\end{widetext}
\end{document}